\documentclass[11pt,a4paper]{article}

\pdfoutput=1

\usepackage{graphicx}
\usepackage{jinstpub}
\usepackage{mathtools}
\usepackage{lineno}

\title{Using machine learning to separate hadronic and electromagnetic interactions in the GlueX forward calorimeter}

\author{Rebecca~Barsotti}
\author{and Matthew~R.~Shepherd}
\affiliation{Department of Physics, Indiana University, Bloomington, IN 47405}
\emailAdd{mashephe@indiana.edu}

 \abstract{
     The GlueX forward calorimeter is an array of 2800 lead glass modules that was constructed to detect photons produced in the decays of hadrons. A background to this process originates from hadronic interactions in the calorimeter, which, in some instances, can be difficult to distinguish from low energy photon interactions.  Machine learning techniques were applied to the classification of particle interactions in the GlueX forward calorimeter. The algorithms were trained on data using decays of the $\omega$\ meson,  which contain both true photons and charged particles that interact with the calorimeter. Algorithms were evaluated on efficiency, rate of false positives, run time, and implementation complexity.  An algorithm that utilizes a multi-layer perceptron neural net was deployed in the GlueX software stack and provides a signal efficiency of 85\% with a background rejection of 60\% for an inclusive $\pi^0$ data sample for an intermediate quality constraint.
     }

\begin{document}

\maketitle

\section{Introduction}

The GlueX experiment in Hall~D at Jefferson Lab seeks to study the light meson spectrum utilizing photoproduction on a proton target.  In particular, GlueX intends to study excited hadrons, many of which will produce photons in their decays, either directly or through the decays of $\eta$ and $\pi^0$.  Typically, the analysis of particular photoproduction reactions benefit from exclusive reconstruction, where one seeks to detect all stable particles produced in the initial collision.  In doing so, backgrounds from other reactions can be rejected by requiring that energy and momentum is conserved.  Broadly speaking, the detectable decay products of a photoproduction reaction consist of charged hadrons and photons.  The former are analyzed using tracking chambers, while the latter are measured, in GlueX detector, with a pair of calorimeters:  the barrel calorimeter (BCAL)~\cite{BCAL} and the forward calorimeter (FCAL).  Photons undergo electromagnetic interactions with material in the calorimeters and tend to deposit all of their energy in a localized ``shower" within the calorimeter.  The cross section for hadron interactions is much smaller and this results in partial energy deposition of hadrons and ``split-off showers" in the calorimeter that are displaced from the primary hadron impact point due to propagation (and subsequent interaction) of neutral secondary particles.   The primary shower produced by a hadron interaction is often easy to identify as one can extrapolate the measured charged particle trajectories to the calorimeter.  However, the displaced split-offs tend to mimic low energy photons.  Misidentifying split-off hadronic showers as low energy photons is problematic as it introduces a combinatoric background and upsets energy-conservation requirements of exclusive reconstruction.  There is no strong discriminator between a true low energy photon and a split-off, and hence, we present a machine learning approach to separate these two classes of showers.  This is a long-standing problem in calorimeter reconstruction, and the idea of using machine learning to address it goes back a couple of decades (see, for example, Ref.~\cite{tom_thesis} and references therein).  In what follows we build on this work by presenting new discriminating variables and discuss the application and performance of a machine learning algorithm for split-off rejection in a different event environment.

The GlueX forward calorimeter is a 2-m diameter circular array of 2800 lead glass modules, each $4~\mathrm{cm}\times4~\mathrm{cm}\times45~\mathrm{cm}$, that was constructed to detect photons produced in the decays of hadrons~\cite{FCAL}. The calorimeter is about 5 meters downstream of the target and is designed to detect photons emitted from the primary interaction with a polar angle that is within 11$^\circ$ of the beam axis.  A photon shower occurs in the forward calorimeter when a photon collides with the lead glass block, creating an electron-positron pair. As they travel through the lead glass, they emit Cherenkov and Bremsstrahlung radiation. The Bremsstrahlung radiation then produces more electron-positron pairs and so on, creating a shower of Cherenkov-radiation-emitting particles in the calorimeter. The Cherenkov radiation is detected by a photomultiplier tube and is proportional to the deposited energy.  (A full description of individual calorimeter modules can be found in Ref.~\cite{FCAL}.)  Showers typically span several blocks, and a reconstruction algorithm~\cite{radphi} is used to group blocks into clusters, each of which is assumed to correspond to an individual photon.  The desire for the algorithm to distinguish between independent closely-spaced true photons competes with the desire to associate a split-off secondary shower from a hadron with the primary interaction point.  Rather than attempt to modify the shower reconstruction algorithm, we choose to examine properties of the reconstructed showers in training a machine-learning algorithm to make the distinction between low energy photons and and split-offs.

The showers in the forward calorimeter can be classified as either true electromagnetic showers, produced by photons, or background originating from charged particles or noise. Throughout this work, we adopt the following definitions for these showers.  Type 0 showers are true photon showers from hadron decays (e.g. a $\pi^0$ decay). Type 1 showers originate from charged particles colliding with the calorimeter, as identified geometrically by tracks in the drift chambers leading to the collision point on the calorimeter. Type 2 showers are all other types of showers, these are dominantly split-offs of a Type 1 shower, but can also be background noise, or other such interactions within the detector.  We focus our effort on the distinction between Type 0 and Type 2 showers.

\begin{figure}
    \centering
    \includegraphics[width=.5\textwidth]{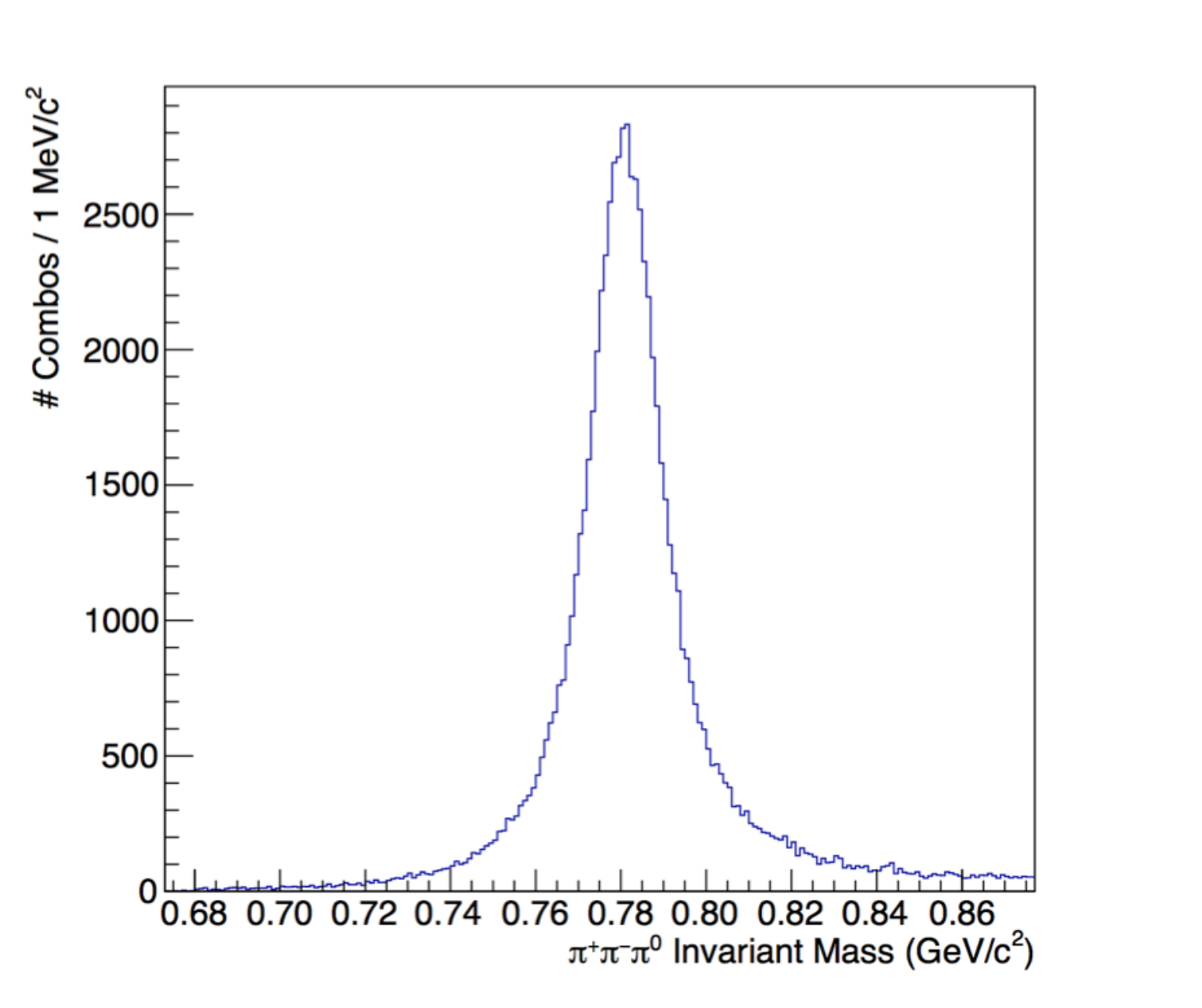}
      \caption{The invariant mass of the $\pi^+\pi^-\pi^0$ candidates used to identify events of the type $\gamma p\to \omega p$.  The peak at 782 MeV/$c^2$ corresponds to signal events.  Very little background exists in the sample.\label{fig:omega}}
\end{figure}

We took a data-driven approach to train and test the algorithms by using exclusively reconstructed events of the type $\gamma p \to \omega p$, where $\omega\to\pi^+\pi^-\pi^0$, a process that has a relatively high cross section and is therefore easy to reconstruct with high purity.  In this reaction the final detected particles are $p\pi^+\pi^-\gamma\gamma$, where the two photons have an invariant mass consistent with the $\pi^0$.  This process produces all types of showers noted above, and allows study of different attributes of particle showers in the calorimeter to be used as distinguishing variables in the machine learning algorithms. In selecting such events, we perform a five-constraint kinematic fit that enforces conservation of four-momentum for both the primary interaction and the $\pi^0$ decay.  We require the confidence level of this fit to be greater than 5\% and the beam energy $E_\gamma$ to be $7.5$~GeV$< E_\gamma < 9.0$~GeV.  Fig.~\ref{fig:omega} shows the $\pi^+\pi^-\pi^0$ invariant mass for events that pass these criteria.  The dominance of $\omega\to\pi^+\pi^-\pi^0$ decays indicates the purity of the sample.  We select events that have a candidate $\omega$ invariant mass in the region of 730-840~MeV/$c^2$.  Within this sample of events, we then classify all showers in the FCAL.  Type 0 showers are those that are used to reconstruct the $\pi^0\to\gamma\gamma$ decay.  Type 1 showers are geometrically matched to the $p,$ $\pi^+$, or $\pi-$ track.  All other showers in the FCAL are classified as Type 2 showers.  The purity of the data sample supports the assumption that the other showers are split-off hadronic interactions or noise.  These samples were then used to study and train eight different types of machine learning algorithms to determine which would give the most effective final classifications between Type 0 and Type 2 showers. Hereafter, we refer to Type 0 as ``signal" and Type 2 as ``background".

\section{Discriminating Variables}

\begin{figure}
    \centering
    \includegraphics[width=.35\textwidth]{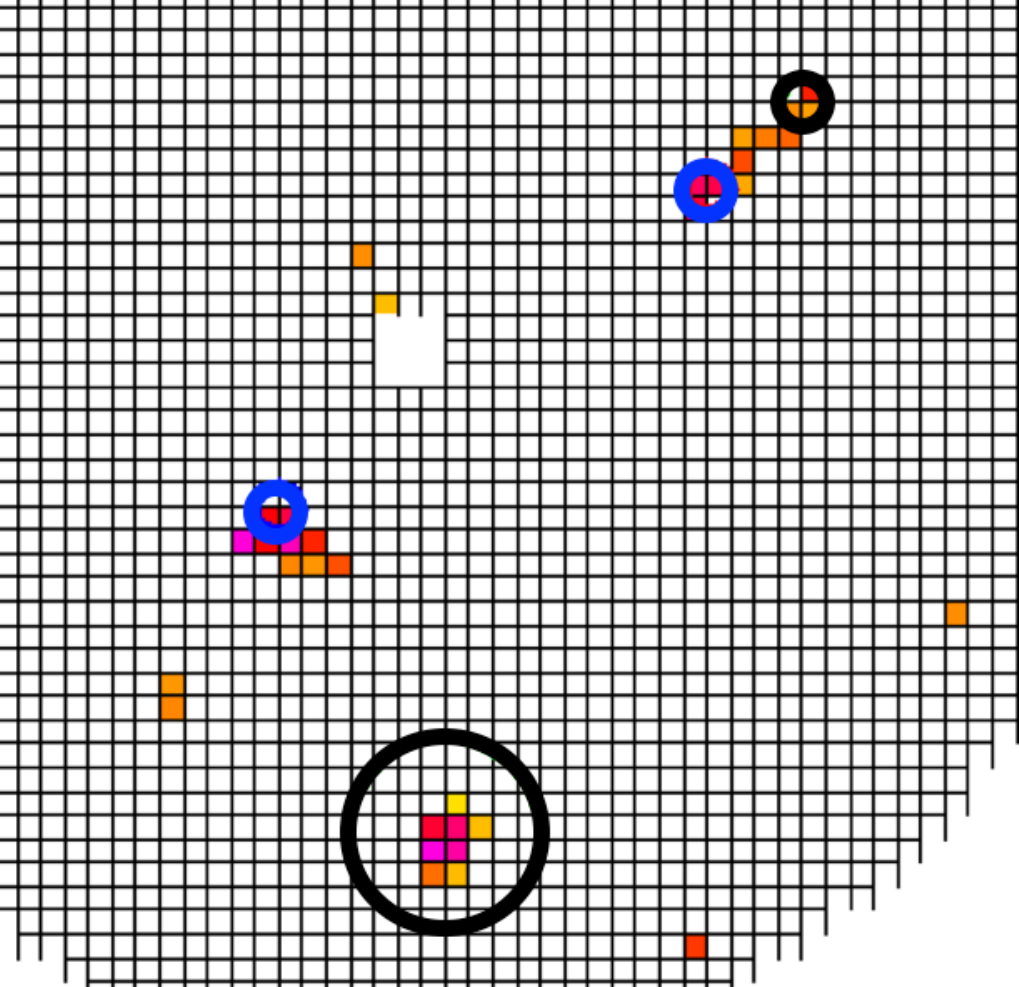}
      \caption{A typical FCAL event. Colored blocks show cells registering an energy deposition. Circles indicate showers identified by the reconstruction algorithm, where the radius of the circle is proportional to the energy of the shower.  Blue circles indicate showers geometrically matched to charged tracks.  Black circles indicate photon candidate showers. The upper right black circle is likely a background shower (hadron split-off) while the bottom black circle is likely a signal shower (true photon).\label{fig:event}}
\end{figure}

Eight variables were selected to train the algorithms based on their ability to differentiate between signal and background showers.  The chosen variables were constructed in order to expose differences in key elements of the two types of showers:  the geometry, the energy distribution, and the timing.  An understanding of the importance of these features can be gained from Fig.~\ref{fig:event}, which illustrates the different spacial distributions of energy between signal and background showers.

Two variables were selected to address the energy distribution of showers: $E_9/E_{25}$ and $E_1/E_9$, where $E_n$ is the total energy deposited in the square array of $n$ modules centered on the module in which the maximum energy was deposited.  These variables tend to be higher for electromagnetic showers than hadronic split-offs because the scale for electromagnetic interactions, where most energy is deposited within the Moli\`ere radius of 2.6~cm, is smaller than nuclear interactions.

\begin{figure}
    \centering
    \includegraphics[width=.5\textwidth]{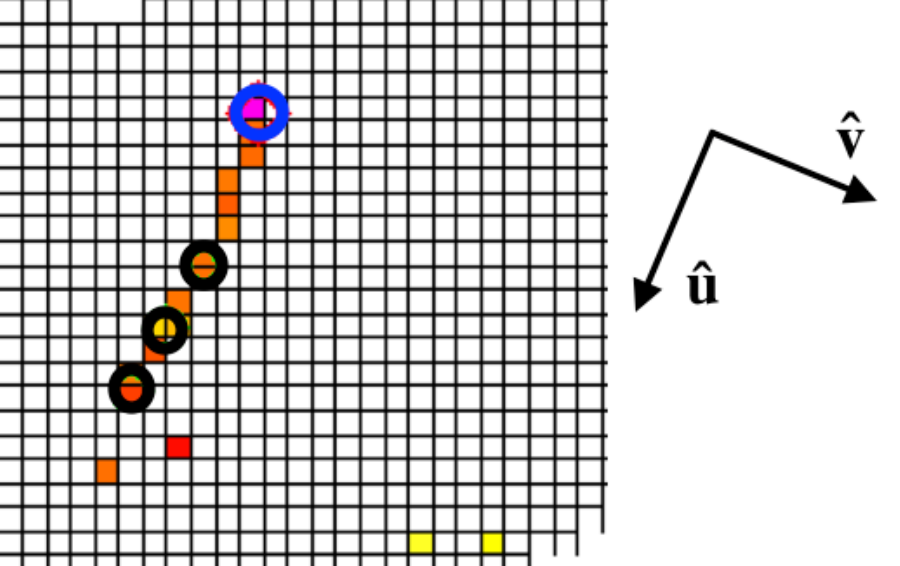}
      \caption{Multiple identified showers in the forward calorimeter originating from a single charged particle.  The blue circle indicates the track-matched shower while the black circles are presumably background showers. The principal axes (ad defined in the text) are sketched.~\label{fig:uv}}
\end{figure}

Fig.~\ref{fig:uv} shows another common type of hadronic split-off.  In this figure it appears that the interaction of a single hadron with the calorimeter (indicated by the blue circle) produced a charged track that propagated through the calorimeter material.  This resulted in the shower identification algorithm finding multiple low-energy photon candidates.  To address such cases, we attempt to construct variables that expose whether a shower has a shape that suggests it came from a nearby track.

We first define the ``principal axes'' of a shower.  If $\mathbf{r}_\mathrm{track}$ is the impact point on the FCAL face of the nearest track to a shower and $\mathbf{r}_\mathrm{shower}$ is the location of the shower on the FCAL face, then we define the unit vectors
\begin{eqnarray}
\hat{\mathbf{u}} &=& \frac{\mathbf{r}_\mathrm{shower}-\mathbf{r}_\mathrm{track}}{|\mathbf{r}_\mathrm{shower}-\mathbf{r}_\mathrm{track}|}, \\
\hat{\mathbf{v}} &=& \hat{\mathbf{u}}\times\hat{\mathbf{z}},
\end{eqnarray}
where $\hat{\mathbf{z}}$ points in the beam direction, normal to the FCAL face.  These principal axes are sketched approximately for the three candidate showers in Fig.~\ref{fig:uv}.  (In reality they vary slightly for each shower as $\mathbf{r}_\mathrm{shower}$ changes.)

With these axes defined, we can construct four variables that discriminate based on the geometry of a shower: $\sigma^2_u$, $\sigma^2_v$, $A_{uv}$ and $N_\mathrm{hits}$. Here $\sigma^2_u$ and $\sigma^2_v$ are the normalized second moments of the energy distribution within a shower about the $u$ or $v$ axes.  For example,
\begin{equation}
\sigma^2_u = \frac{ \sum_{i=1}^{N_\mathrm{hits}} E^\mathrm{hit}_i \left( ( \mathbf{r}^\mathrm{hit}_i - \mathbf{r}_\mathrm{shower}) \cdot \hat{\mathbf{u}}\right)^2}{\sum_{i=1}^{N_\mathrm{hits}} E^\mathrm{hit}_i},
\end{equation}
where $E^\mathrm{hit}_i$ and $\mathbf{r}^\mathrm{hit}_i$ are the energy deposition and module locations of the individual blocks (``hits") that are clustered together to form the shower, $N_{\mathrm{hits}}$ is the number of calorimeter blocks registering an energy deposition in a shower. Showers that are produced from hadronic interactions, tend to be elongated along the $u$ axis.  For true photon showers, the choice of $u$ and $v$ axes is effectively random as it depends on the random location of the nearest track and hence there is no difference between $\sigma^2_u$ and $\sigma^2_v$ for true photons.

In order to expose the asymmetric nature of hadronic split-offs, we define an asymmetry variable:
\begin{equation}
    A_{uv} = \left | \frac{ \sigma^2_u - \sigma^2_v }{\sigma^2_u + \sigma^2_v} \right |,
\end{equation}
which tends toward one for hadronic split-offs but toward zero for true photons.  It is anticipated that the
precision of these variables will depend on the number of blocks in the shower being considered; therefore,
we include $N_\mathrm{hits}$ as an input to the machine learning algorithm also.

Lastly, two variables discriminate based on the timing of the showers: $c_\mathrm{eff}$ and $\Delta t$. We define $c_\mathrm{eff}$ as an effective velocity given by the distance from the interaction point to the shower divided by the difference in time at the interaction point and the shower. The variable $\Delta t$ is the difference between the time of shower and the time of impact closest track to the shower. Details about the timing resolution of the calorimeter can be found in Ref.~\cite{FCAL}.  The distributions of $c_\mathrm{eff}$ and $\Delta t$ are slightly different for signal and background showers; however, the behavior of these variables is not intuitive.  For example, the distribution of $c_\mathrm{eff}$ for background showers tends towards values greater than the speed of light.  These effects are attributed to differences in the propagation time of Cherenkov photons through the lead glass bar for these two types of showers, which depends on the point of origin of these photons within the block.  Cherenkov photons from signal showers tends to be emitted furthest from the photomultiplier tube at an angle such that they bounce numerous times before reaching the downstream end of the block.

\begin{figure*}
  \centering
  \includegraphics[width =0.9\textwidth]{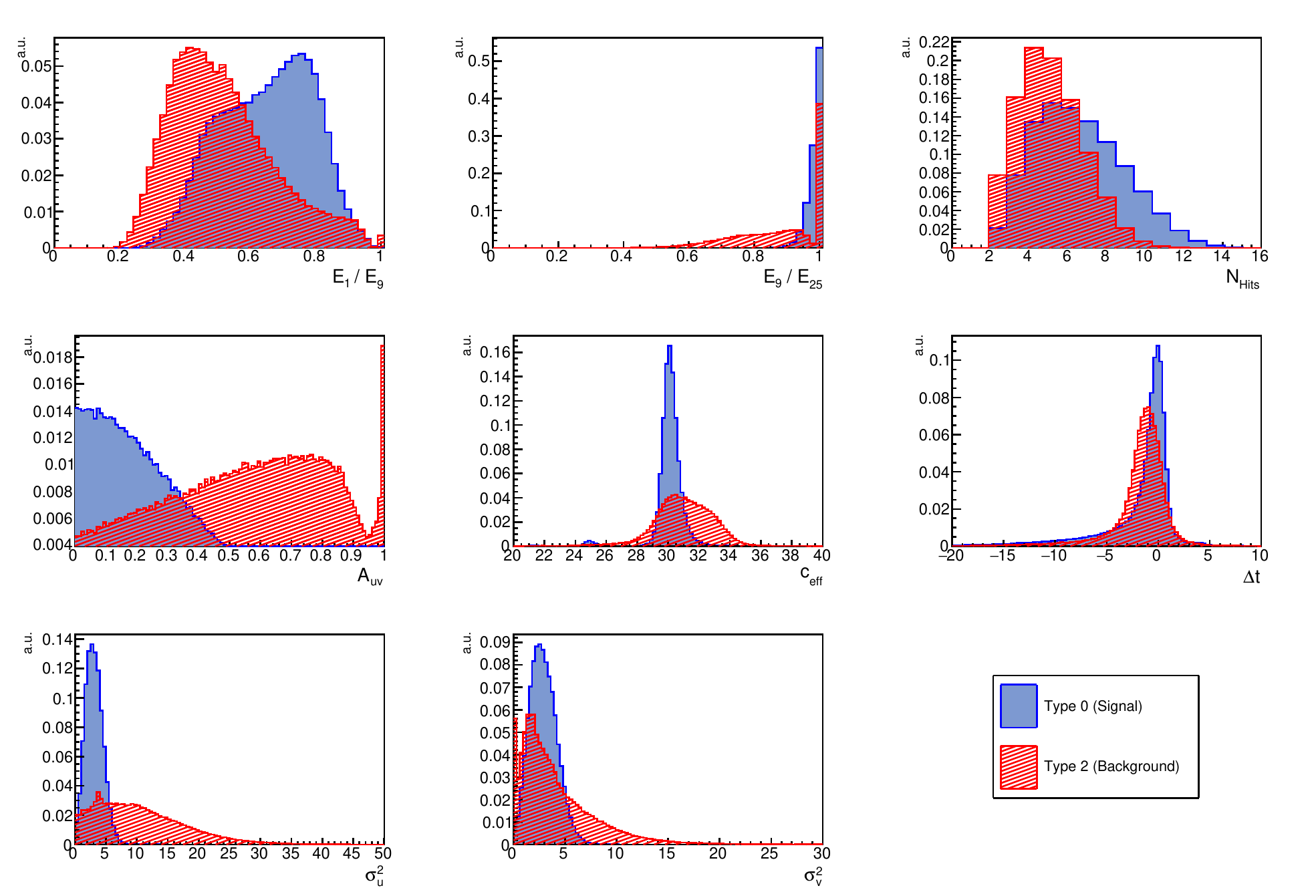}
  \caption{Input variables to the machine learning algorithms. Red histograms represent the distribution for background showers, while the blue histograms represent the distributions for signal showers.  The plots were generated using data from exclusively reconstructed events, as discussed in the text.\label{fig:vars}}
\end{figure*}

The complete set of input variables, derived from events of the type $\gamma p \to \omega p$, for signal and background showers is shown in Fig.~\ref{fig:vars}. 

An important consideration in the selection of these variables, apart from their ability to discriminate between shower types, was their process independence. Variables like the distance to the nearest track-matched shower and the total energy of the shower, while likely useful, were excluded due to concern that the algorithm may perform in a significantly different way for different event topologies. For example, in high track-density areas of the calorimeter, distance to nearest track-matched shower could be misleading.  Also one would expect the energy of the shower to depend on the process that produced the particles.

\begin{figure}
\begin{center}
    \includegraphics[width=0.45\textwidth]{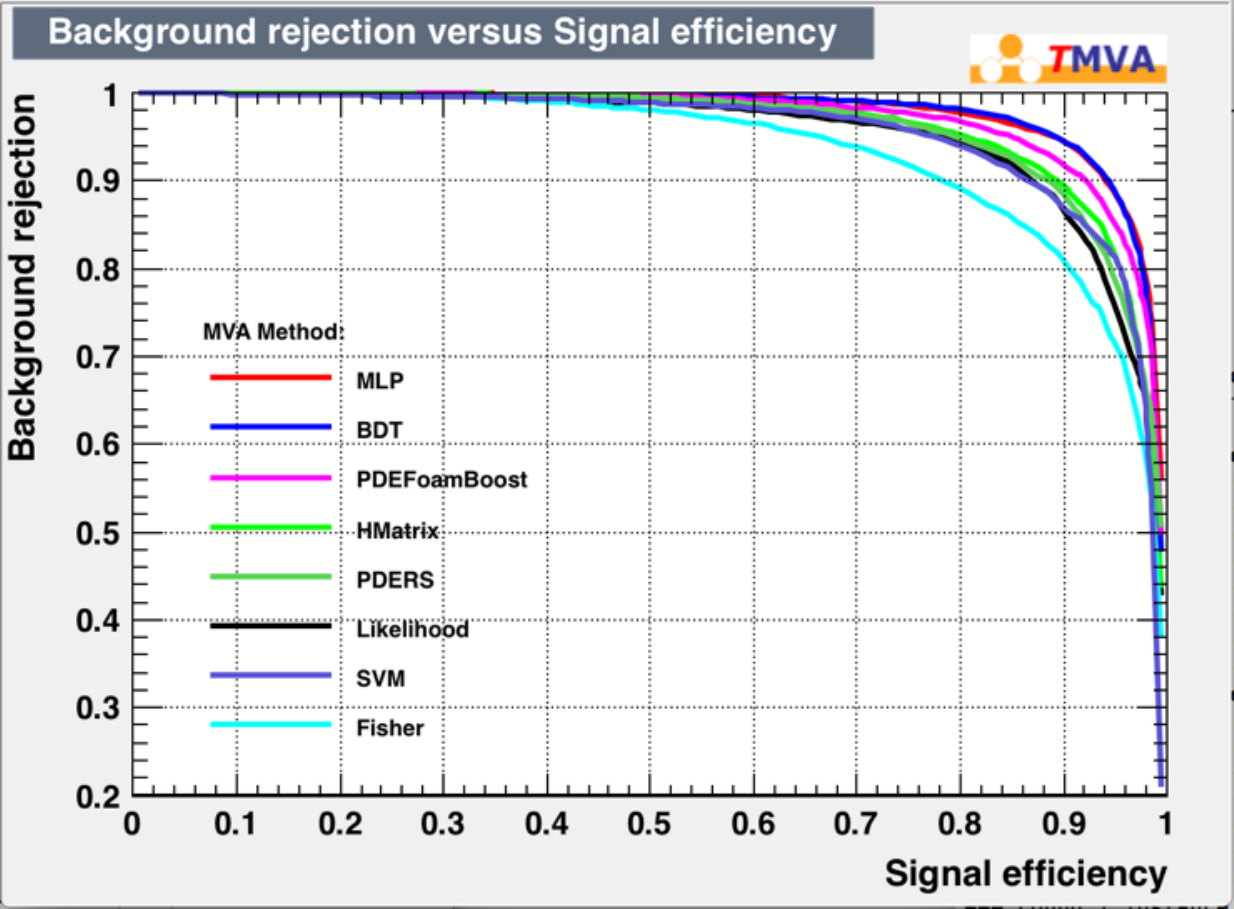}
    \includegraphics[width=0.5\textwidth]{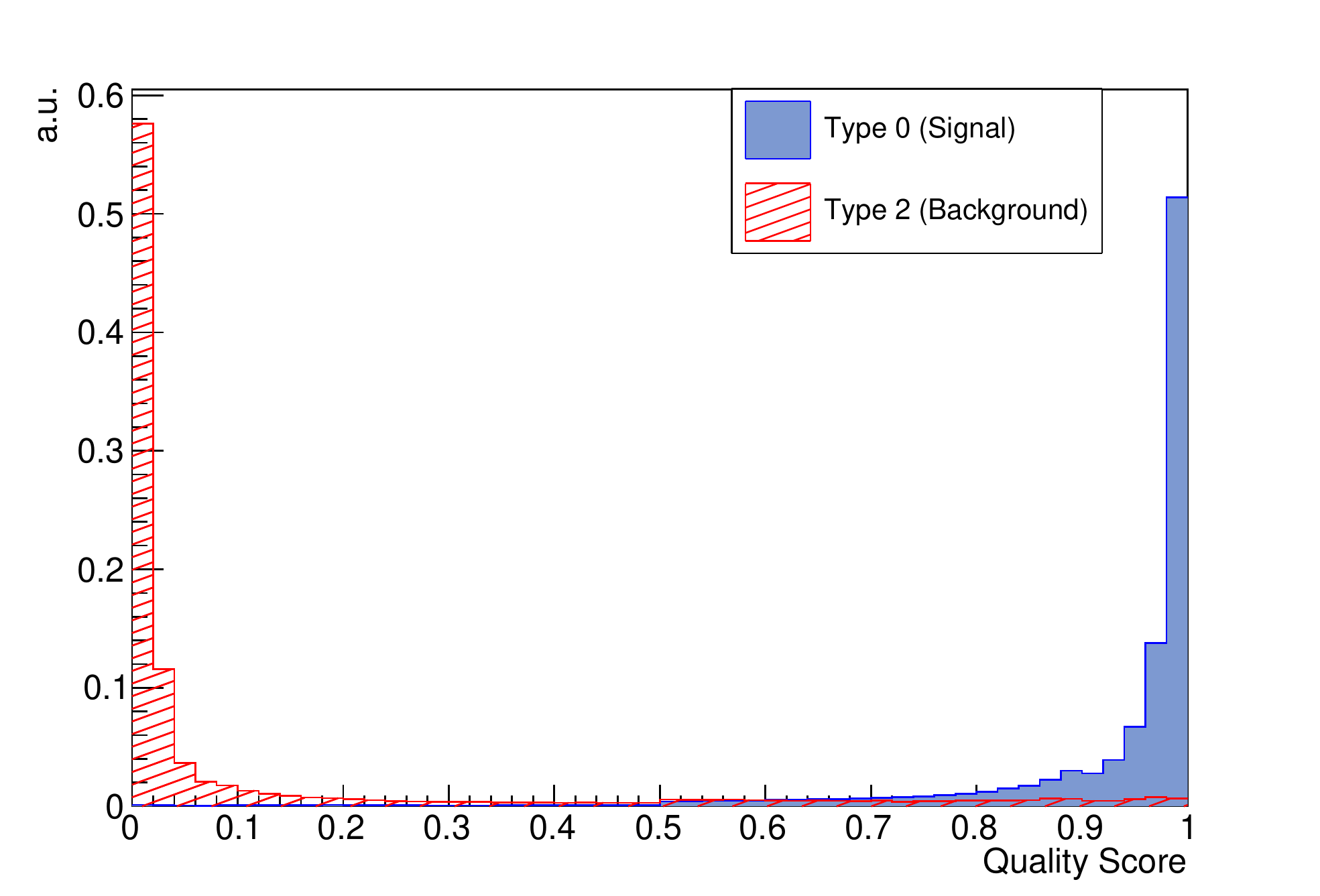}
    \caption{(top) performance of the eight different types of multi-variate analysis algorithms that were tested. (bottom) classifier output distribution for the MLP when run on the background (red) and signal (blue) training samples.\label{fig:algs}}
    \end{center}
\end{figure}

The eight selected variables were used to train and test eight different machine learning algorithms in the Toolkit for Multivariate Analysis (TMVA)~\cite{TMVA2007}. Fig.~\ref{fig:algs} shows the background rejection vs. signal efficiency curves for these eight algorithms.  The top performing algorithms tested were the multi-layer perceptron (MLP) and the boosted decision tree (BDT), with nearly identical performance. Ultimately, the MLP was selected due to ease of implementation within the GlueX software framework.  The MLP can be simply coded as a standalone function of the input variables, whereas the BDT requires storing and reading the trees that are used in the boosted decision tree algorithm.
The classification output of the MLP algorithm for the training samples is shown in Fig.~\ref{fig:algs}.  It is evident that placing a requirement on this quantity provides much better discrimination between signal and background showers than any of the variables in Fig.~\ref{fig:vars}.
 
\section{Algorithm performance}

A number of tests were executed on the chosen MLP algorithm to quantify the performance on generic photoproduction events. For these performance tests, we inclusively selected events that contained two photons whose invariant mass were within 50~MeV/$c^2$ of the $\pi^0$ mass.  This sample, which is representative of a general event environment, was then used to study the purity of $\pi^0$ reconstruction. The neural network algorithm functions by assigning a single quality score between zero and one to each photon candidate, with one being extremely photon-like and zero very unlike a photon. Photon candidates that are geometrically matched to the track trajectories extrapolated to the calori{-}meter are assigned a quality score of zero.  In addition, we find that stable fits to the $\pi^0$ peak can only be reliably obtained in all interesting regimes to study if we exclude  photon candidates with a quality score less than 0.05.  Doing so results in a negligible efficiency loss and provides a stable ``denominator" for probing the performance of the algorithm with this $\pi^0$ sample.  In the studies that follow we only consider the effects of enforcing minimum quality requirements greater than 0.05.

\subsection{Quantifying performance}

\begin{figure}
    \centering
    \includegraphics[width=0.5\textwidth]{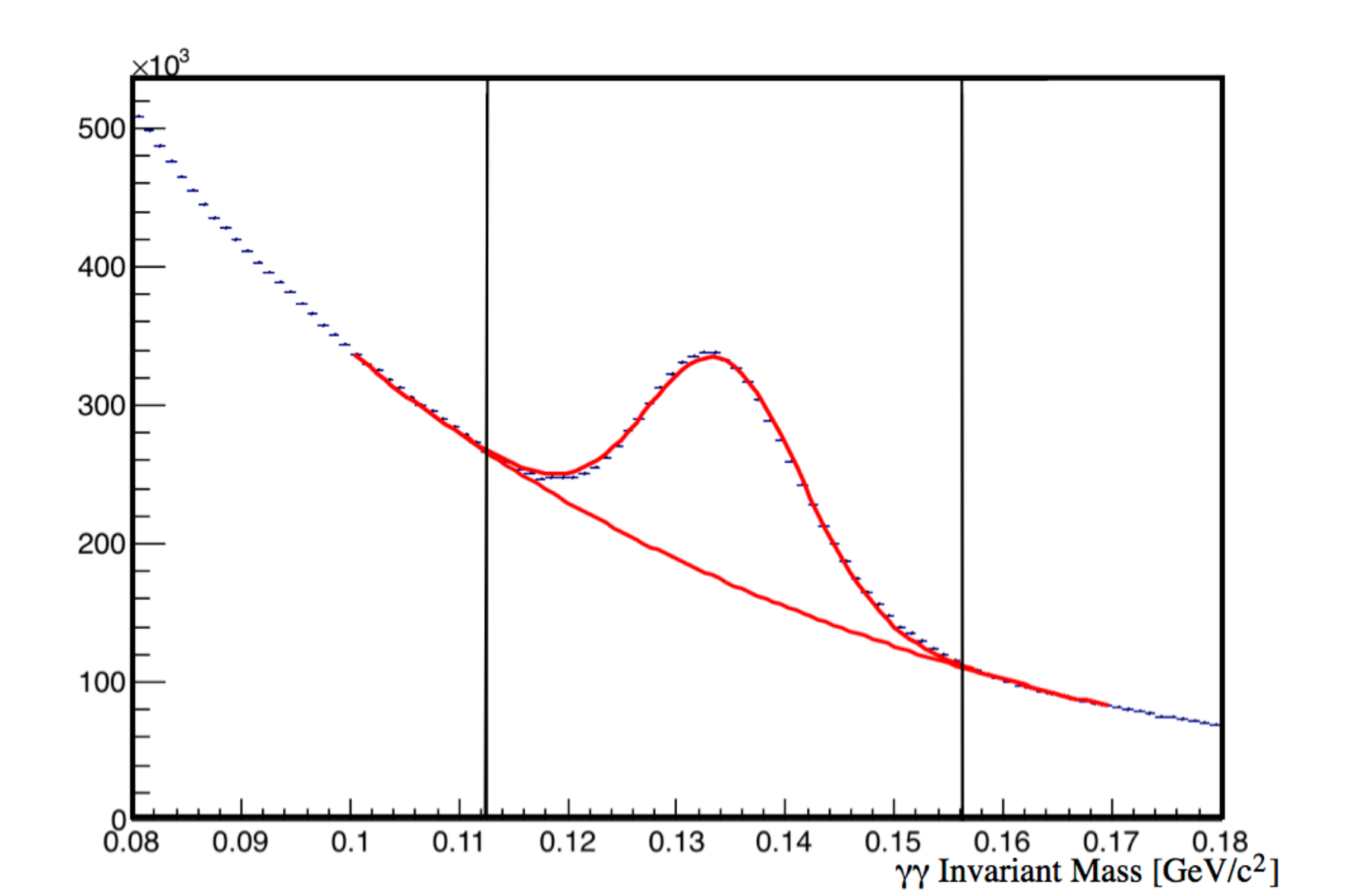}
      \caption{Invariant mass of two photons in the $\pi^{0}$ sample with fitted signal Gaussian and polynomial background distributions. Vertical lines indicate the signal region.\label{fig:pi0}}
\end{figure}

In order study the performance of the algorithm, it is useful to define construct several quantities that quantify the features of the $\pi^0$ invariant mass distribution.  Figure~\ref{fig:pi0} shows a typical two-photon invariant mass distribution.  We fit a signal Gaussian and polynomial background function to the spectrum.  The quantities $S$ and $B$ are the integrals of these functions within the signal region, defined as $\pm3\sigma$ of the mean.  While the background may contain some true photons that did not come from $\pi^0$ decays, the $\pi^0$ signal is only composed of true photons.  Therefore, any reduction of $\pi^0$ signal by placing a requirement on the minimum quality score of the photon is an indication of an inefficiency in the algorithm.  Because the background is only partially from split-off showers, the background rejection effectiveness of the algorithm is likely dependent on the specific topology one tries to reconstruct.

With these definitions of $S$ and $B$, we can then construct several useful quantities.  One commonly used quantity is a figure of merit proportional to the statistical significance of the $\pi^0$ signal
\begin{equation}
\mathrm{FOM} = \frac{S}{\sqrt{S+B}}.
\end{equation}
where $S$ denotes the signal yield and $B$ denotes the background yield. We may also construct the signal to background ratio $S/B$.  Finally, we can construct relative efficiency as a function of the minimum quality constraint $q_\mathrm{min}$
\begin{equation}
    \mathrm{RE}(q_\mathrm{min}) = \frac{ S_{\mathrm{quality}>q_\mathrm{min}}}{ S_{\mathrm{quality}>0.05}},
    \label{eq:re}
\end{equation}
where, in the denominator we have applied, as discussed above, a minimum quality requirement of 0.05 to facilitate stable fitting of signal function to the spectra.

\subsection{Performance in different event environments}

The simplest method for exploring how the algorithm performs in different environments is to sort events based on the number of reconstructed tracks.  Increasing the track multiplicity increases the number of split-off showers in the calorimeter, and hence sorting events on this quantity is likely to expose various trends in performance.

\begin{figure}
\begin{center}
    \includegraphics[width=.49\textwidth]{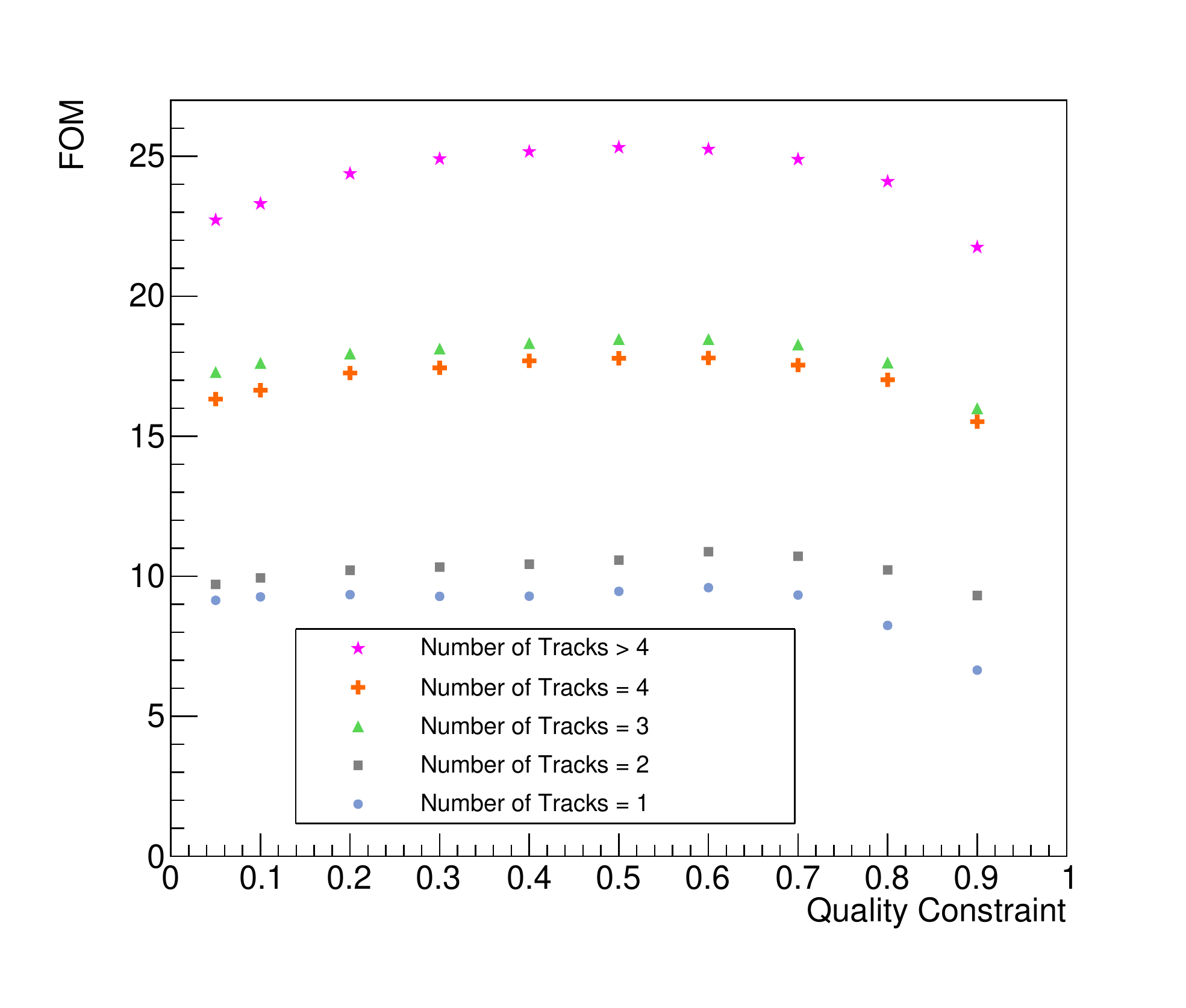}
    \includegraphics[width=.49\textwidth]{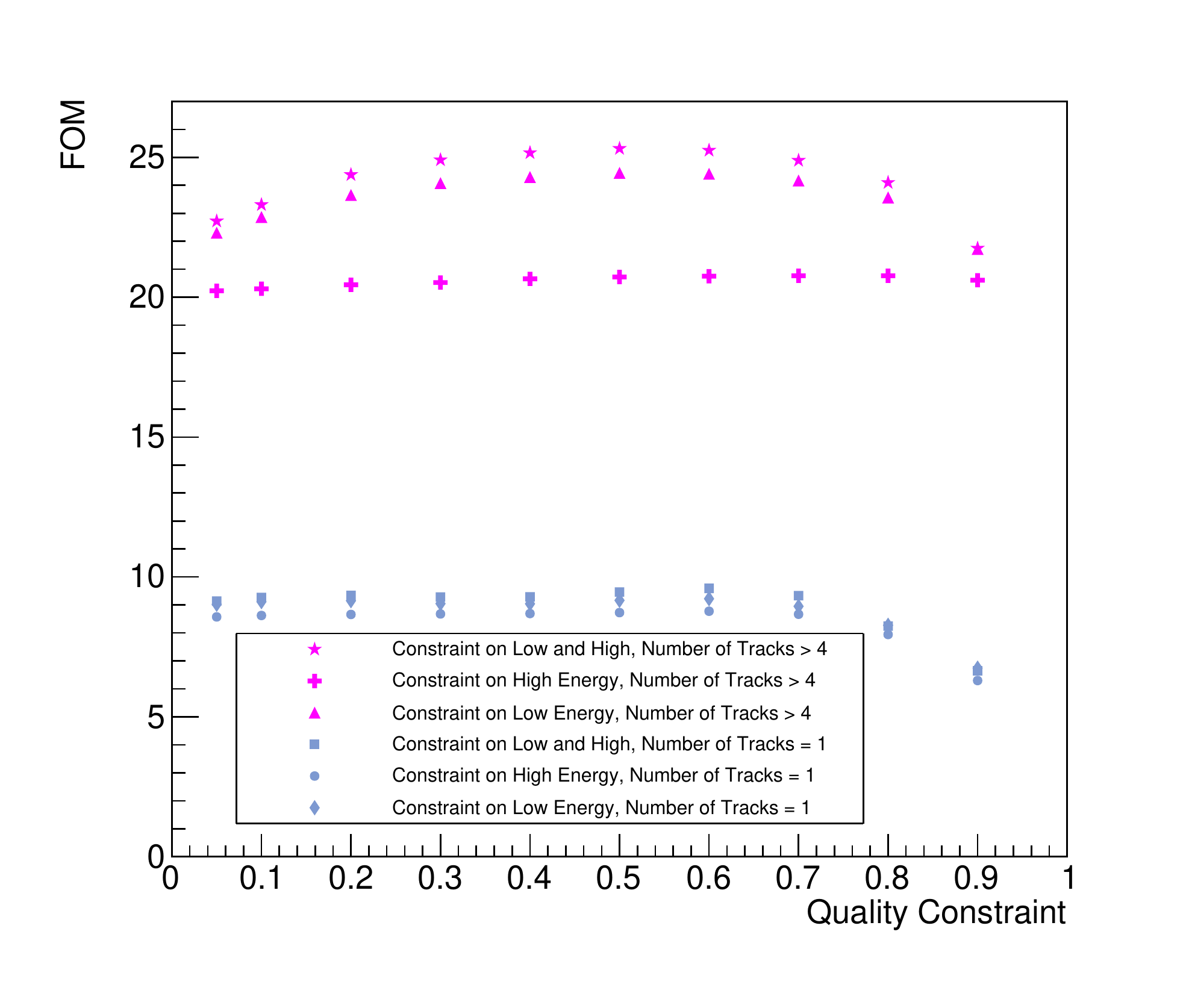}
    \caption{The figure of merit studies broken up by number of tracks in an event (left) and by constraining the quality of only the higher energy photon, only the lower energy photon, and both simultaneously (right). \label{fig:ntrack}}
\end{center}
\end{figure}

Figure~\ref{fig:ntrack} shows the FOM vs. minimum quality requirement for different numbers of tracks in the event.  There are a few key features to note in this plot. At first glance, it may appear that there is little gain in the FOM when constraining the quality score.  One must recall, that ensuring fit stability requires the first point corresponds to a constraint of quality greater than 0.05.  The greatest increase in the FOM occurs when placing a constraint to be any higher than zero, which is not shown in this plot.  Events with larger numbers of tracks tend to have a higher FOM, but this simply due to variations in the overall number of $\pi^0$'s in the event samples and not a feature of the algorithm. The key feature of this plot is the relative change in the FOM as a function of quality score and not the absolute scale of the FOM.  More importantly, Fig.~\ref{fig:ntrack} indicates that all types of events have a broad maximum between quality constraints of 0.3 and 0.7, which suggests that a common requirement in this range is suitable for maximizing signal significance.

The utility of placing separate minimum quality requirements on the photons from the $\pi^0$ decay is studied separately.  Figure~\ref{fig:ntrack} shows the FOM vs.\ quality constraint when applied to the high-energy photon, low-energy photon, or both.  We show this for extreme cases in number of tracks:  events with only one track and events with more than four tracks.  The largest gains in FOM are obtained when quality requirements are applied to the low-energy photon in events with a high track multiplicity.  Placing additional requirements on the high-energy photon provides some incremental improvement that is only noticeable in events with large numbers of tracks.  In all other studies in this note, the quality constraint is applied to both the low-energy and high-energy photons.

\begin{figure}
\begin{center}
    \includegraphics[width=.49\textwidth]{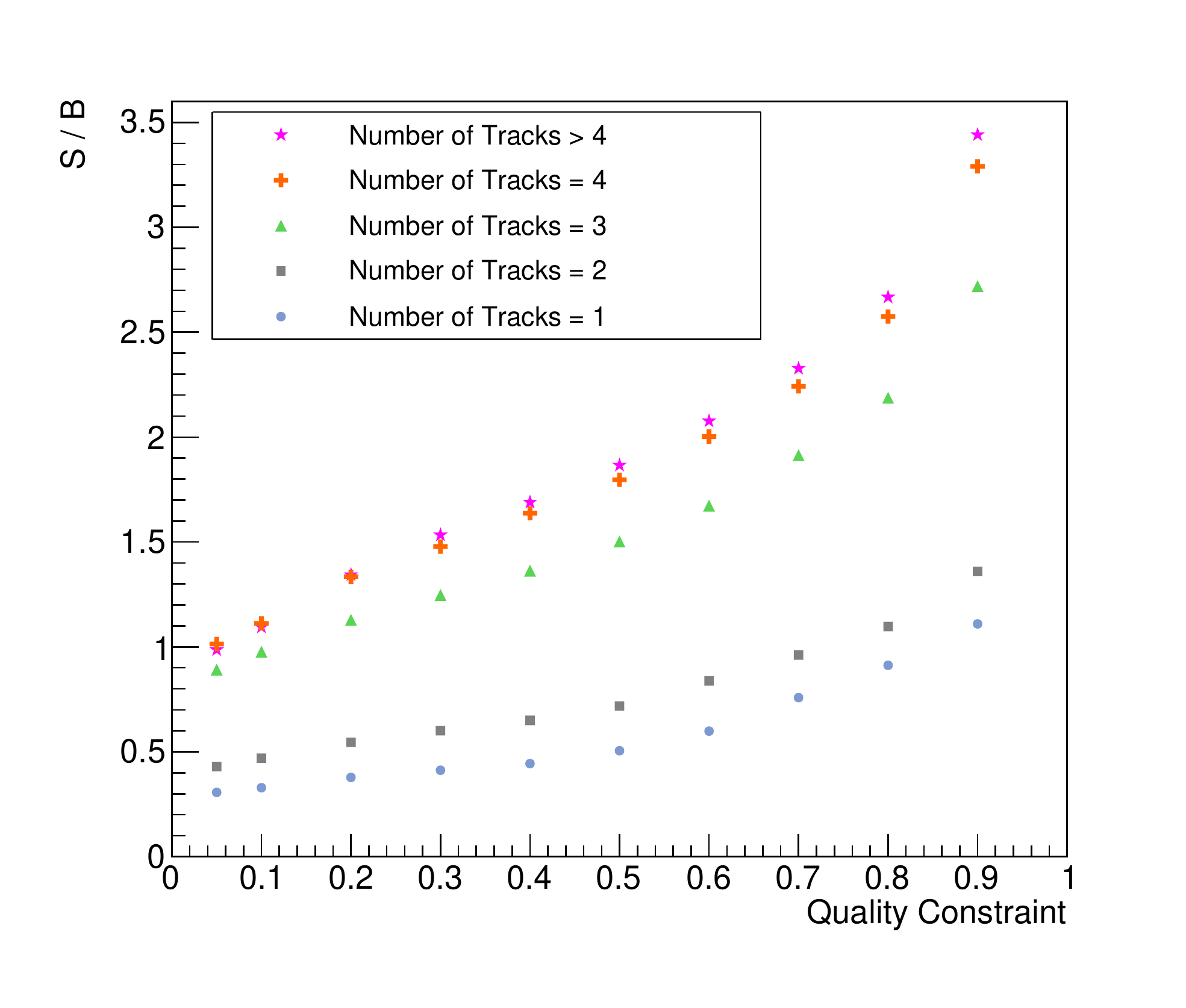}
    \includegraphics[width=.49\textwidth]{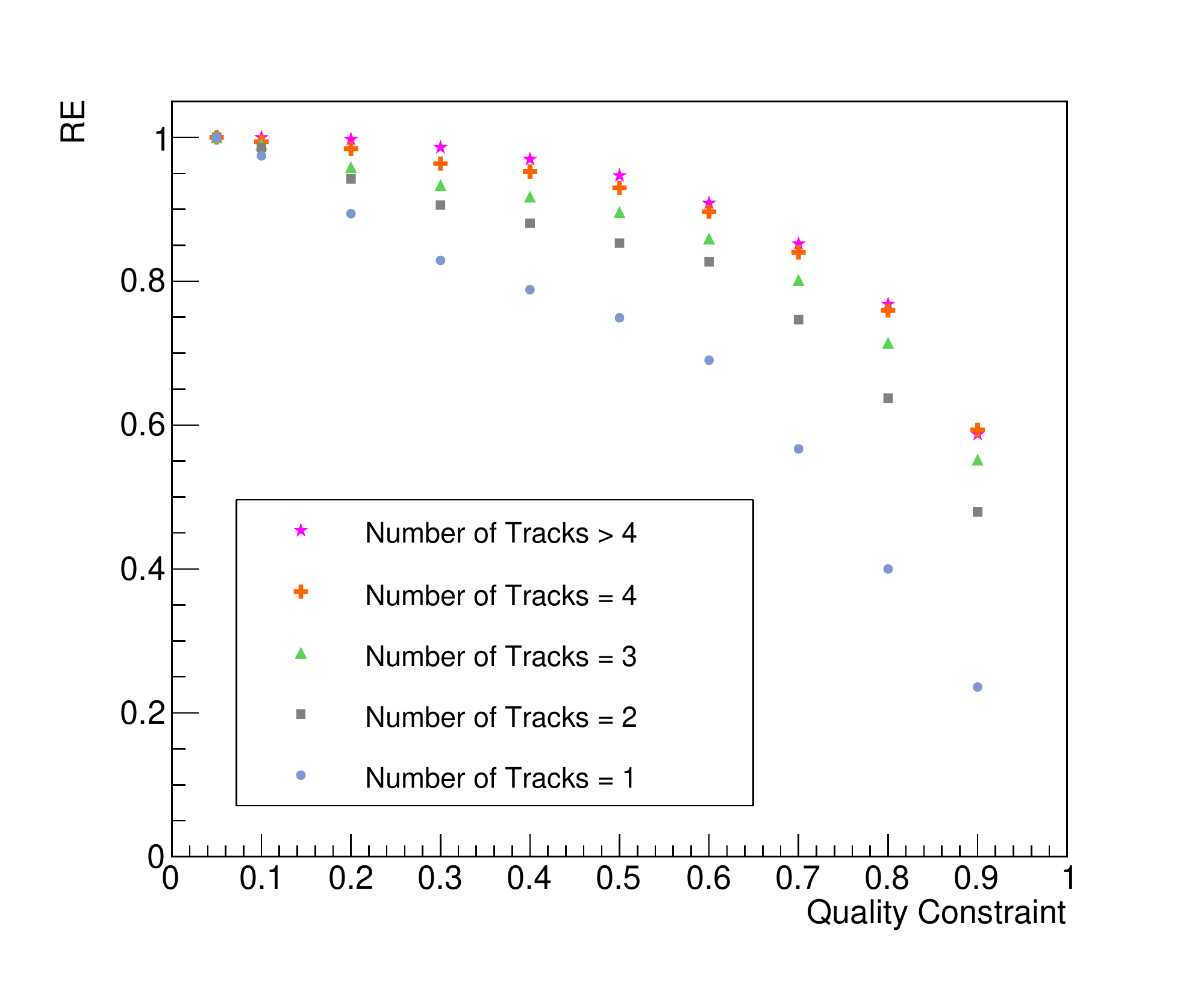}
    \caption{Signal to background ratio (top) and relative efficiency (bottom) for different track numbers by quality requirement. \label{fig:releff}}
\end{center}
\end{figure}

Finally in Fig.~\ref{fig:releff}, we plot the signal-to-background ratio $S/B$ as well as the relative efficiency RE (defined in Eq.~\ref{eq:re}) for varying numbers of tracks in the event.  One can see that the greatest improvements in $S/B$ are obtained in events with large numbers of tracks.  In addition, the quality requirements are typically more inefficient for events with low numbers of tracks, which may be due to the fact that the event vertex, and hence timing variables related it, is poorly defined in one-track events. Although any quality requirement increases $S/B$ even for events with one track.

\subsection{Simulation of efficiency}

In order for the algorithm to be practically applicable in data analysis, it is important that the \textsc{Geant4}-based~\cite{geant4} Monte Carlo (MC) simulation of the detector produces a detector response such that the efficiency of the selection algorithm in MC matches that of data. Any deviation will result in a systematic error in photon reconstruction efficiency in MC.  With the relative efficiency as defined in Eq.~\ref{eq:re}, we compare inclusive data and MC simulated samples of $\pi^0$ events as well as data and MC simulated samples of exclusive $\omega$ production.

\begin{figure}
 \begin{center}
    \includegraphics[width=0.49\textwidth]{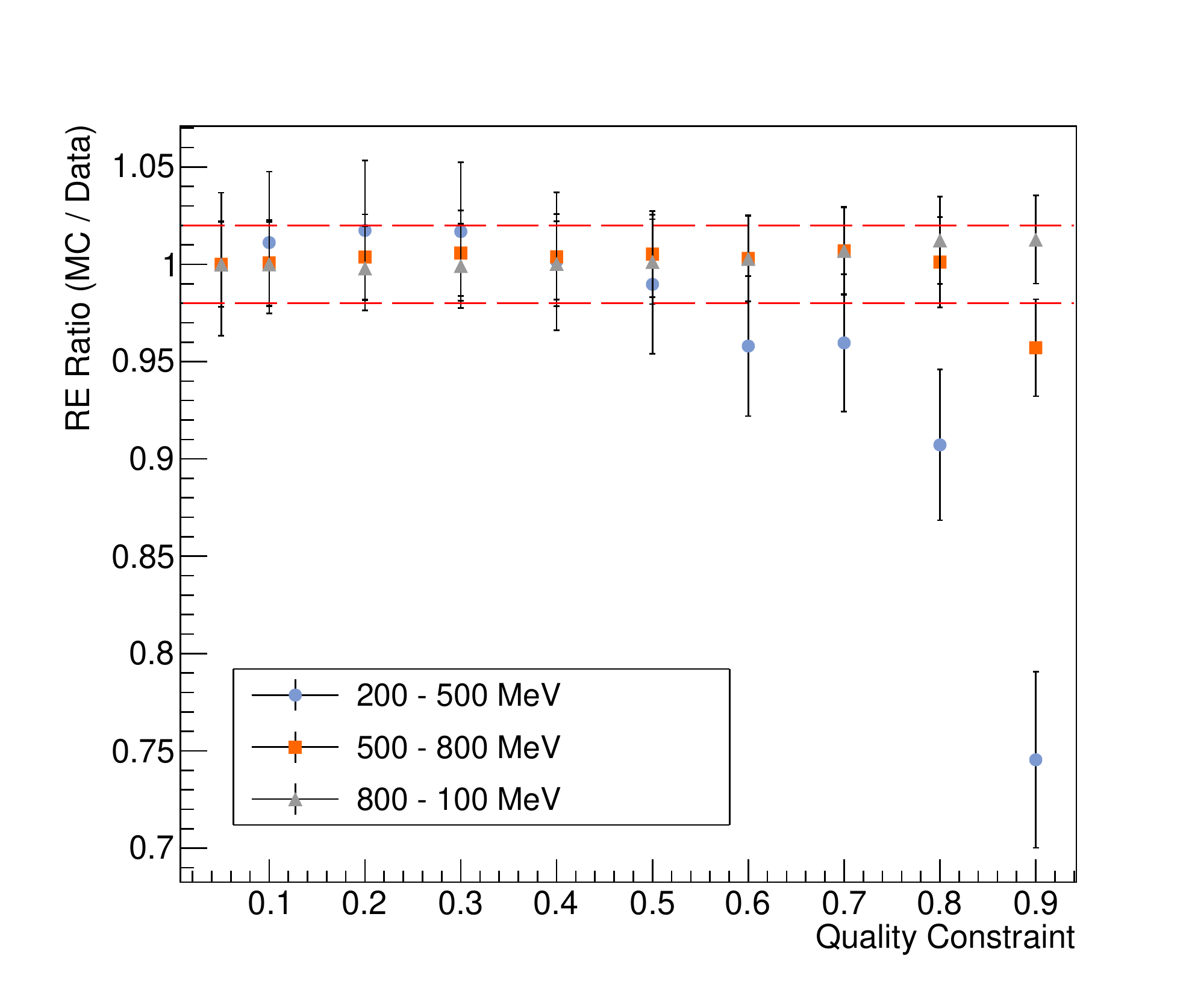}
    \includegraphics[width=0.49\textwidth]{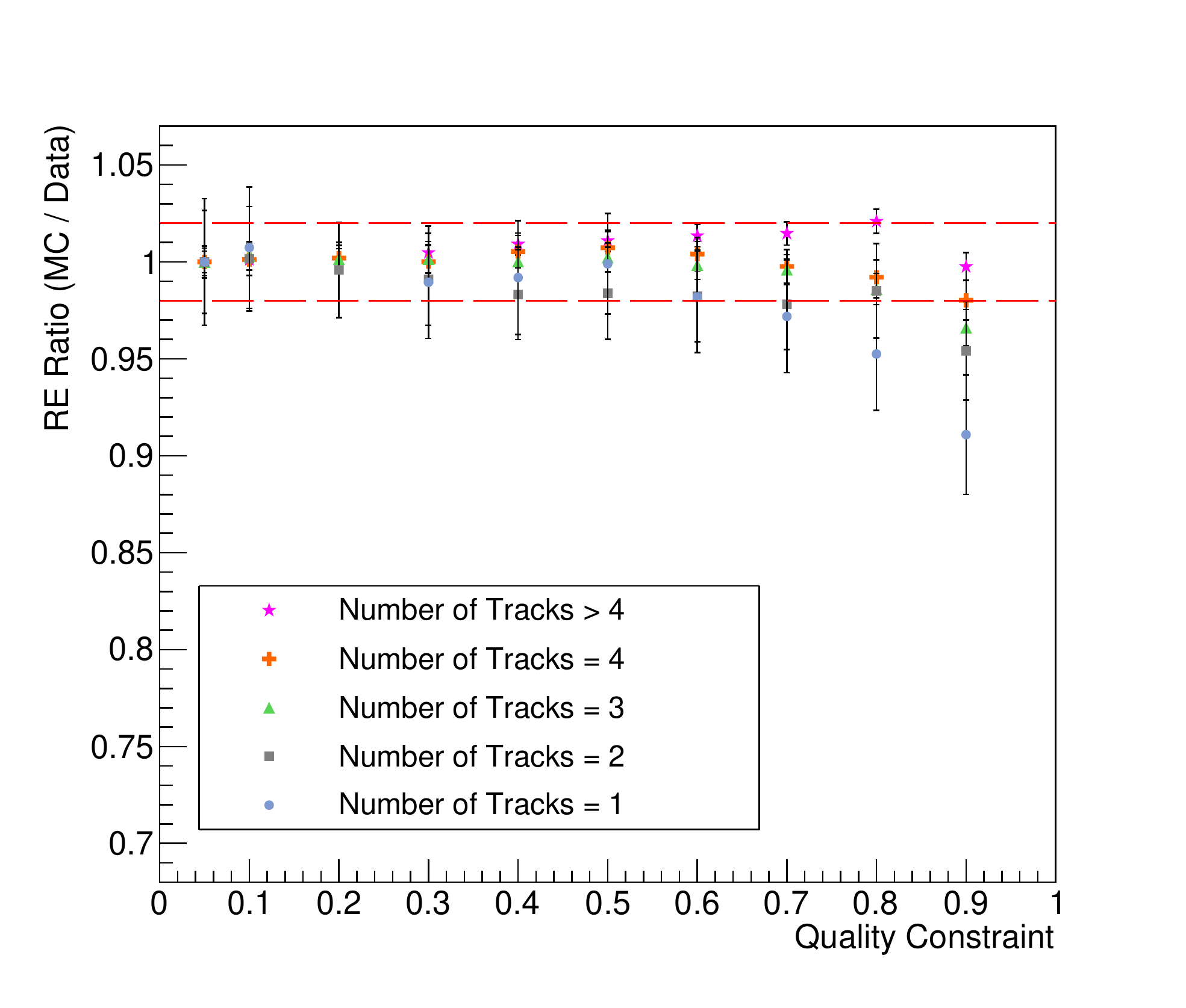}
      \caption{Monte Carlo and Data efficiency ratio for various energy ranges and event environments. Dashed lines indicate +/- 2\%. (left) from exclusive $\gamma p \to \omega p$ events in both MC and data.  The higher energy photon of the $\pi^0$ pair fixed with a quality constraint of 0.5, while the lower energy photon has a varying quality constraint. (right) from inclusive $\pi^0$ events in MC and data.  The low-energy photon is between 500~MeV and 800~MeV and high-energy photon above 1.0 GeV with varying quality constraint imposed on both. \label{fig:MCData}}
  \end{center}

\end{figure}

In the first part of this study, we aim to study how well the MC models the efficiency dependence on energy in a common event environment.  We examine exclusive $\gamma p \to \omega p$, $\omega \to \pi^+\pi^-\pi^0$ samples using the low-energy photon of the $\pi^0$ decay as our probe.  The high-energy photon is required to have a minimum quality score of $0.5$ and we vary the quality constraint of the low energy photon.  The energy of the low-energy photon is also used to sort the sample into three regions.  Figure~\ref{fig:MCData} shows the efficiency ratios as a function of quality constraint on the low-energy photon for the three energy regions.  One sees agreement at the few percent level over the entire phase space except for the most stringent quality constraints at low energy, where there are a large number of split-off showers.

In the second part of this study, we look at the inclusive $\pi^0$ samples in data and MC to examine how well the MC reproduces the efficiency dependence on event environment for photons in some range of energy. The $\pi^0$ events were taken from an inclusive data sample and an MC that models inclusive photoproduction of hadrons. We require that the low-energy photon in the $\pi^0$ pair is between 500~MeV and 800~MeV and that the high-energy photon is above 1.0 GeV. We impose the same quality constraint on both photons and sort the sample by the number of tracks in the event. Figure~\ref{fig:MCData} shows the efficiency ratios as a function of quality constraint on both photons for the various numbers of tracks in the event. Based on these results, and the previous studies, we conclude that a minimum quality requirement of 0.5 maximizes FOM, and the efficiency of this requirement is simulated by MC at the few percent level.

\subsection{Typical application}

\begin{figure}
\begin{center}
    \includegraphics[width=.49\textwidth]{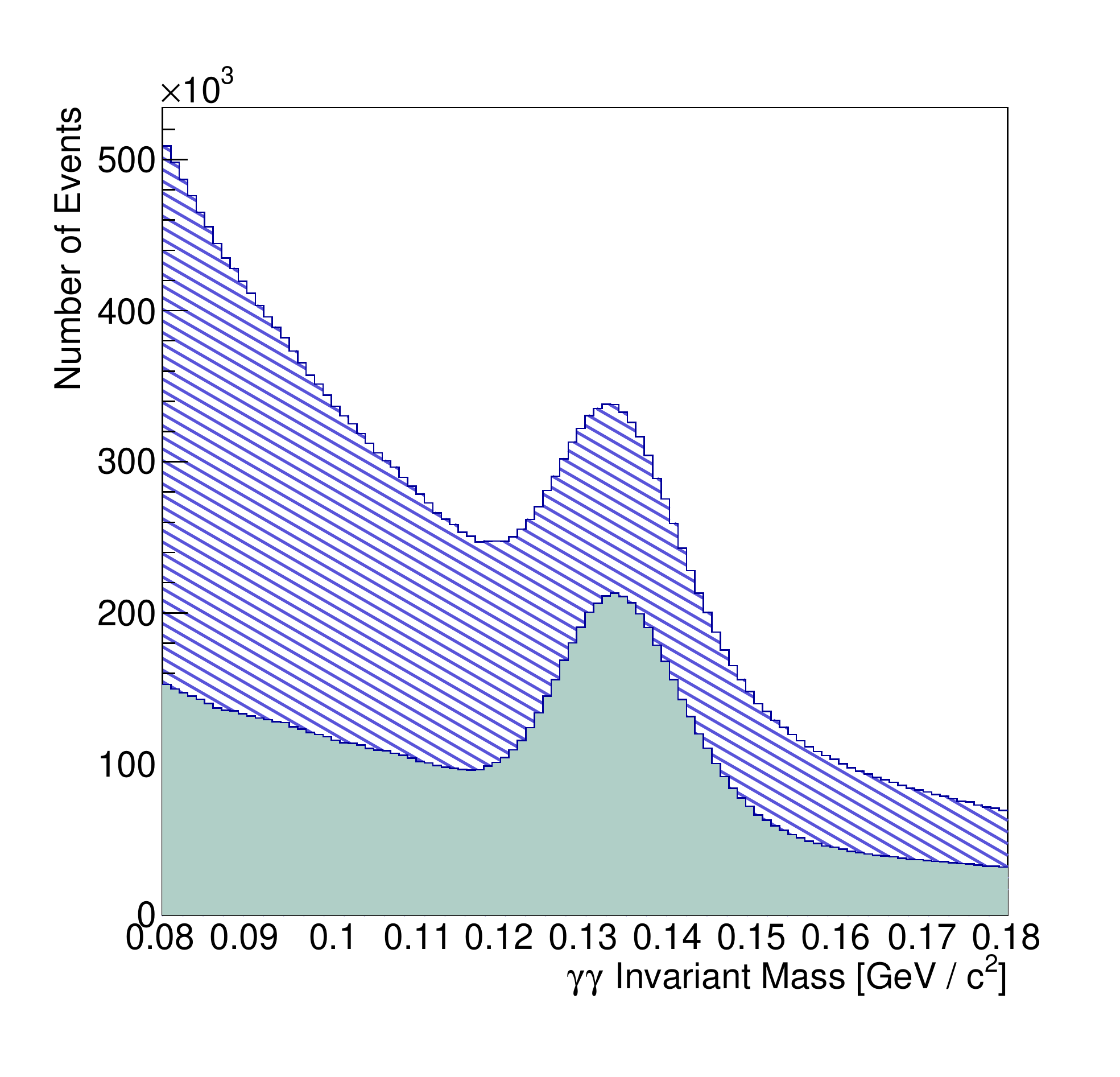}
    \includegraphics[width=.49\textwidth]{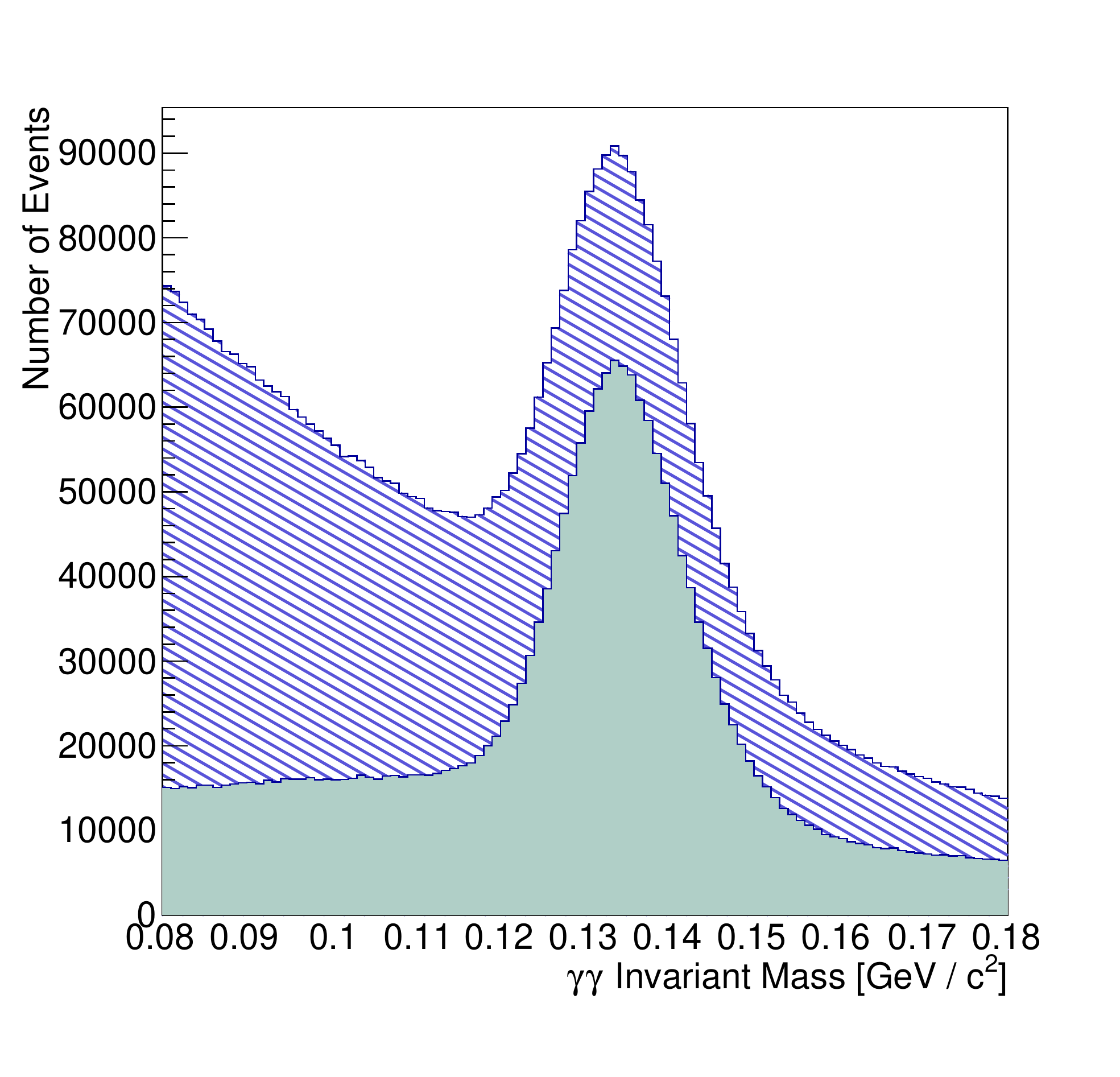}
    \caption{Reduction of combinatoric background resulting from a quality requirement of 0.5. Data with no quality requirement is in striped blue, and data with the quality requirement is in solid green. All number of tracks (left), data with four or more tracks (right). \label{fig:bkgreduct}}
\end{center}
\end{figure}

Based on the studies above, we apply a quality requirement of 0.5 to both photons from a $\pi^0$ decay and examine the performance.  Figure~\ref{fig:bkgreduct} shows spectra before and after application of our algorithm and requirement.  In this figure, we show as a starting point the spectra obtained by only discarding showers that are geometrically matched to tracks; therefore, the background reduction that is visible reflects the full gain obtained by the algorithm.  For all events, a minimum quality requirement of 0.5 yields a background reduction of roughly 60\%, and retains roughly 85\% of the signal. For events with greater than four tracks, this requirement provides a background reduction of 64\% while retaining 96\% of the signal. 

Imposing stricter quality requirements leads to a significant reduction in background, but at the cost of some of the signal peak and increased systematic uncertainty due to MC modeling of the efficiency. Requiring photon quality to be higher than 0.7 on all events, for instance, yields a background reduction of roughly 77\% and retains about 75\% of the signal. The strictest constraint tested was requiring quality greater than 0.9. On events where the number of tracks exceeded four, this constraint yielded a 87\% background reduction but retained only 60\% of the signal. 

\section{Summary}
The use of machine learning techniques to reduce background in the GlueX forward calorimeter is a powerful tool. Using defining characteristics of the energy depositions in the calorimeter blocks, such as the shape, size, and distribution, we were able to train a good discriminating neural network. The optimal quality requirement for the figure of merit yields a background reduction of 60\% and a signal retention of 85\% on inclusive $\pi^0$ data. The Monte Carlo and data efficiency for this requirement, when studied with $\omega$ events, agree within the available statistical precision.

\acknowledgments

We would like to acknowledge support of our colleagues in the GlueX collaboration who facilitated collection and processing of data as well as development of the simulation of the GlueX detector.  In particular we would like to thank R.~Jones and M.~Dalton for their helpful comments and suggestions on this study and manuscript.  This work was supported by the US Department of Energy Office of Nuclear Physics under grant DE-FG-02-05ER41374, and the NSF REU program under grant PHY-1757646.

\end{document}